\documentclass
[aps,pre,showpacs,floatfix,showkeys,superscriptaddress,twocolumn]{revtex4}
\usepackage{amssymb}
\usepackage{amsmath}
\usepackage{amsfonts}
\usepackage{graphicx}
\usepackage{epsfig}

\def\be{\begin{equation}}
\def\ee{\end{equation}}
\def\beq{\begin{eqnarray}}
\def\eeq{\end{eqnarray}}

\begin{document}
\title{Exponentially enhanced quantum metrology}

\author{S.\ M.\ Roy}
\affiliation{Computer Science, University of York,
York YO 10 5DD, United Kingdom}
\affiliation{Department of Theoretical Physics, Tata Institute of
Fundamental Research,\\ Homi Bhabha Road, Mumbai 400005, India}
\author{Samuel L.\ Braunstein}
\affiliation{Computer Science, University of York,
York YO 10 5DD, United Kingdom}

\date{\today}

\begin{abstract}
We show that when a suitable entanglement generating unitary
operator depending on a parameter is applied on $N$ qubits in
parallel, a precision of order $2^{-N}$ in estimating the parameter
may be achieved. This exponentially improves the precision
achievable in classical and in quantum non-entangling strategies.



\end{abstract}

\maketitle

The Mandelstam-Tamm version of the
time energy uncertainty relation \cite{ref1} and its rigorous developments
\cite{ref2} form the basis of quantum enhanced methods for parameter
estimation such as those used in evolving frequency standards \cite{ref3}.
Giovannetti, Lloyd and Maccone \cite{ref4} have recently done beautiful
work to introduce a general framework to classify methods of such quantum
enhancement. A unitary transformation depending on the parameter to
be estimated is applied to a suitably prepared system of $N$ probes
and then an appropriate observable is measured. A separable initial
state is labelled C and an entangled state Q. Similarly, measured
observables which are direct products of individual qubit
observables are designated C and those which introduce entanglement
are designated Q. Thus the strategies of ``parallel'' measurement on
$N$ qubits are classified as CC, CQ, QC, QQ. Their result is that
quantum metrology in the strategies CC and CQ only achieve the
classical-precision limit of order $\epsilon /\sqrt N$ , where
$\epsilon$ is the dispersion of measurement results for each probe;
but the quantum metrology strategies QC and QQ can achieve a
precision of order $\epsilon /N$ if the probes are prepared in a
suitable entangled state. In contrast, for $N$ sequential
measurements on a single probe to achieve the same precision of
order $1/N$ requires a running time and hence duration of quantum
coherence to be $N$ times longer.

\begin{figure}[ht]
\begin {center}
\includegraphics[width=6.5cm,height= 2.5cm,angle=0]{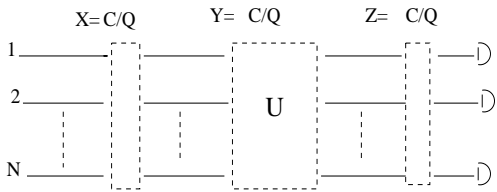}
\end {center}
\caption{The eight parallel strategies $XYZ$ for estimation of a
parameter occurring in an unitary operator $U$ applied to $N$ probes
classified according to types of state preparation,unitary
operation, and measurement. Each of $X,Y,Z$ takes values $C$ or $Q$.
$Y=Q$ if $U$ is an entanglement generating operator (as considered
in this paper) and $Y=C$ otherwise. $X,Z$ specify presence of
entanglement ($Q$) or absence of entanglement ($C$) in the prepared
probe state and in the measured operator respectively.  }
\label{Figure 1}
\end{figure}

We propose here a parallel strategy which exploits the Hilbert space
of $N$ probes more fully than in previous work \cite{ref4} and
thereby attains an exponentially enhanced precision. In discussing
parallel strategies previous work have only considered applying on
the prepared probes a unitary operator which is a direct product of
$N$ unitary operators each acting on a single probe. We will show
here that if instead we consider applying on the probes, an
entanglement generating unitary operator $U = e^{- i\theta H}$ which
{\it cannot\/} be written as a direct product of one-probe
operators, then we can obtain an exponentially enhanced precision in
estimating the parameter $\theta$. The fundamental reason for this
improvement is that there are an exponentially large number of
mutually commuting observables for the $N$-probe system whereas the
number of mutually commuting single-probe operators is only of order
$N$. For instance, for the $N$-qubit system each qubit cannot have
more than one commuting observable and hence there are exactly $N$
commuting observables of the form ${\openone}\otimes \cdots \otimes
{\openone}\otimes A^{j} \otimes{\openone}\otimes\ldots{\openone} $
each of which acts non-trivially on only the $j$th qubit, with
$j=1,2,\ldots,N$. We construct sets of $2^{N-1}$ mutually commuting
Hermitian operators each of the form $A^{1}\otimes A^{2}\otimes
\cdots\otimes A^{N} $ which acts non-trivially on all the $N$
qubits. The unitary entanglement generating operator is chosen to be
a product (but not a direct product) of the exponentials of these
Hermitian operators times $-i\theta $. In this way, we fully exploit
the quantum parallelism which is at the heart of exponential
violations of local realism \cite{ref5} and of the well-known
exponential speed-up achieved in certain quantum computation tasks
\cite{ref6}. For the probe system of $N$ qubits we obtain the best
possible precision of order $ \epsilon\,2^{-N} $ dictated by the
uncertainty principle for such entanglement-generating unitary
operators. The previous precision limit of  order $\epsilon/N$ is
the best possible for unitary operators which are direct products.
We thus obtain a new characterization of eight parallel quantum
metrology strategies as $XYZ$ where each of $X$, $Y$ and $Z$ can be
Q or C. See Fig.~\ref{Figure 1}. Here $X$ and $Z$ specify presence
or absence of entanglement in the probes and observables
respectively as in \cite{ref4}, and the new label $Y$ specifies
whether the unitary operator $U$ applied on the probes is an
entanglement-generating operator or a direct product.

In our parallel strategy, a precision of order $\epsilon\, 2^{-N} $
is obtained using an entanglement generating operator that is a
product of $2^{N-1}$ mutually commuting unitary operators, but
entanglement at the probe preparation stage and measurement stage
are inessential. To obtain the same precision in the sequential
strategy of Ref.~\onlinecite{ref4} one needs an exponentially large
number of applications of a unitary operator on a single qubit;
hence the time for which quantum coherence has to be maintained will
be longer by a factor of order $2^{N}$. The increase in precision we
achieve with respect to the parallel strategy of
Ref.~\onlinecite{ref4} is however at the cost of a ``Hamiltonian''
or ``generator'' of $U$ which might be more difficult to implement
experimentally. Here again, quantum mechanics just might come to the
rescue. We propose a quantum-optics model of laser light interacting
with an $N$-qubit system, say a polyatomic molecule, via a
generalized Jaynes-Cummings interaction which, in principle, could
achieve the exponentially enhanced precision. The practical
implementability of this specific model is at the moment unclear,
but would be an interesting subject for further theoretical and
experimental investigation.

\vskip 0.2truein
\noindent\textbf{Quantum limits on precision}: We recall first the
Mandelstam-Tamm uncertainty relations. Suppose we apply an unitary
operator $U = e^{- i\theta H}$ to a probe state $|\phi\rangle$ where
$H$ is a self-adjoint operator and $\theta$ a parameter to be
estimated. We obtain the state
\be |\psi (\theta)\rangle = U |\phi\rangle\;,
\ee
and then measure the observable $X$ on this state with a view to
estimating the parameter $\theta$. Schwarz inequality yields
\beq
 \Delta X \Delta H &\geq&
|\langle\psi(\theta)|[H,X]/(2i)|\psi(\theta)\rangle|\nonumber\\
& = & \frac{1}{2} \left|\frac{\partial \langle\psi(\theta)|X
|\psi(\theta)\rangle} {\partial \theta}\right| \;,
\eeq
where $\Delta X$, $\Delta H$ denote the dispersions in $X$, $H$
respectively. The resulting precision in estimating $\theta$ is thus
given by the uncertainty relation
\be
\delta\theta \equiv \Delta X /
\left|\frac{\partial \langle \psi
(\theta)|X|\psi (\theta)\rangle} { \partial \theta}\right|
\geq\frac{1}{2\Delta H}\;.
 \ee

In the case when the parameter $\theta$ is estimated using an
estimator  $\theta_{\rm est}$ and the estimation is repeated $\nu$
times, the Cramer-Rao bound \cite{ref7} was used in
Ref.~\onlinecite{ref2} to prove the generalized uncertainty relation
for the error estimate \be \delta_{\nu}\theta
\equiv\bigg<\bigg(\theta_{\rm est}\bigg/
\biggl|\frac{\partial\langle\theta_{\rm est}\rangle_{\rm
av}}{\partial \theta}\biggr|-\theta \bigg)^{2}\bigg>_{\rm av}^{1/2}
\geq\frac{1}{2\Delta H \sqrt \nu} \;, \ee where
$\langle\cdots\rangle_{\rm av}$ denotes statistical average. The
quantum limits on precision are obtained by noting that \be \Delta H
\leq \frac{1}{2}(\lambda_{\rm max}-\lambda_{\rm min}) \;, \ee where
$\lambda_{\rm max}$, $\lambda_{\rm min}$ denote respectively the
maximum and minimum eigenvalues of $H$ which we assume to be finite.
Hence, the Mandelstam-Tamm \cite{ref1} and Braunstein-Caves
\cite{ref2} quantum limits on precision are given respectively by
\be \delta\theta \geq\frac{1}{(\lambda_{\rm max}-\lambda_{\rm min})}
\;, \ee and \be \delta_{\nu}\theta \geq\frac{1}{(\lambda_{\rm
max}-\lambda_{\rm min})\sqrt \nu} \;. \ee It is clear that quantum
parallel strategies to improve precision should aim to maximize
$\Delta H$ on the $N$-probe quantum state.

\vskip 0.2truein
\noindent\textbf{Parallel strategies for $N$ qubits}: Consider
first, as in \cite{ref4} the operator $H$ to be a direct sum of the
operators $H^{j}$ acting on the $j$-th probe, each $H^{j}$ having the
same maximum dispersion $\Delta H^{j}\leq d$. Then
 \be
\Delta \bigoplus_{j=1}^{N} H^{j} \leq N d \;.
\ee
As noted in Ref.~\onlinecite{ref4} this
maximum dispersion is in fact reached when we choose the $N$-qubit
state to be an equally weighted superposition of the eigenvectors of
$H$ with maximum and minimum eigenvalues. With  $\lambda_{\rm max}= N
\lambda_{M}$ and $\lambda_{\rm min}= N\lambda_{m}$ where $\lambda_{M}$
and $\lambda_{m}$ are respectively the maximum and minimum
eigenvalues of each $H^{j}$ we get the above equation with $d=
(\lambda_{M}-\lambda_{m})/2$. Note for purposes of comparison, that
the dispersion of a sum of $K$ classical variables $H^{j}_{\rm cl}$ with
a factorized joint probability distribution is given by
\be
\Delta \sum^{K}_{j=1}H^{j}_{\rm cl}=\sqrt{\sum^{K}_{j=1} (\Delta
H^{j}_{\rm cl})^{2} }\;.
\ee

The maximum quantum dispersion is thus $\sqrt{N}$ times the
classical value when $K=N$ and $\Delta H^{j}_{\rm cl}= d$. This has been
exploited in Ref.~\onlinecite{ref4}. We now show that the dispersion
of $H$ for
$N$ qubits can be made exponentially larger by appropriate choice of
$H$. Consider, the operator identity
\be
\bigotimes _{j=1} ^{N}\bigl(\sigma_{x} + i \sigma_{y}\bigr)^{j}=
H + i A\;,
\label{eq10}
\ee
where $\sigma_{x}^{j}$ and $\sigma_{y}^{j}$ are Pauli
matrices for the $j$-th qubit, and $H$ and $i A$ denote respectively
the Hermitian and anti-Hermitian parts of the operator on the
left-hand side of the equation. Explicit expressions for $H$ and $A$
are conveniently stated in terms of the matrices $\sigma^{j}(\pm 1)$
defined by
 \be
\sigma^{j}(+1)\equiv \sigma_{x}^{j}\;,\qquad \sigma^{j}(-1)\equiv
 \sigma_{y}^{j} \;.
\ee
We obtain
\be
H= \!\!
\sum _{\genfrac{}{}{0pt}{}{r_{1},r_{2},\ldots,r_{N}=\pm 1}{ N_{-}={\rm even}} }
\!\! H(r_{1},r_{2},\ldots,r_{N})\;,
\ee
where
\be
H(r_{1},r_{2},\ldots,r_{N})\equiv (-1)^{N_{-}/2}\bigotimes _{j=1}
^{N}\sigma^{j}(r_{j})\;,
\ee
and
 \be
A= \!\!\sum
_{\genfrac{}{}{0pt}{}{r_{1},r_{2},\ldots,r_{N}=\pm 1}{N_{-}={\rm odd}} }
\!\!A(r_{1},r_{2},\ldots,r_{N})\;,
\ee
where
\be
A(r_{1},r_{2},\ldots,r_{N})\equiv (-1)^{(N_{-}-1)/2}\bigotimes
_{j=1} ^{N}\sigma^{j}(r_{j})\;,
\ee
and finally
 \be
N_{-}=\sum_{j=1}^{N} \frac{1}{2} (1-r_{j})\;,
\ee
is just the
number of $j$'s with $r_{j}=-1$ or the number of
$\sigma_{y}^{j}$'s in the $N$-fold product of Pauli matrices in
$H$ and $A$. Both $H$ and $A$ are sums of $2^{N-1}$ products of
Pauli matrices, each product having eigenvalues $\pm 1$ and hence
maximum dispersion
 \be
\Delta \bigotimes _{j=1} ^{N}\sigma^{j}(r_{j})\leq 1 \;.
\ee

The standard anti-commutation rules between Pauli matrices leads to
\be
\sigma^{j}(r_{j})
\sigma^{j}(r_{j}') =\sigma^{j}(r_{j}') \sigma^{j}(r_{j})r_{j}r_{j}'
\;,
\ee
for $r_{j}, r_{j}' =\pm 1$. Note that $r_{1}r_{2}\ldots
r_{N}=(-1)^{N_{-}}$. The anti-commutation rules then imply that the
set of $2^{N-1}$ products of Pauli matrices occurring in $H$ (or $A$)
constitutes a set of mutually commuting observables. Hence
\be
e^{-i\theta H}=\!\!
\prod_{\genfrac{}{}{0pt}{}{r_{1},r_{2},\ldots,r_{N}=\pm 1}{N_{-}={\rm even}}}
\!\!e^{-i\theta H (r_{1},r_{2},\ldots,r_{N} )}\;,
\ee
and
\be
 e^{-i\theta
A}=\!\!
\prod_{\genfrac{}{}{0pt}{}{r_{1},r_{2},\ldots,r_{N}=\pm 1}{N_{-}={\rm odd}} }
\!\! e^{-i\theta A (r_{1},r_{2},\ldots,r_{N} )}\;.
\ee

Further any of these $2^{N-1}$ observables in $H$ anti-commutes with
any of the $2^{N-1}$ observables occurring in $A$. In
contrast with the parallel strategy in Ref.~\onlinecite{ref4}, we now have
\be
\Delta H \leq  2^{N-1}\;,\qquad \Delta A \leq  2^{N-1}\;.
\ee
Interestingly, and in contrast with Ref.~\onlinecite{ref4} the
maximum dispersions of $H$ and $A$ are now reached in the
separable $N$-qubit states
\be
|\!\uparrow \uparrow \ldots \uparrow \uparrow\rangle \;,\qquad
|\!\downarrow \downarrow \ldots \downarrow \downarrow \rangle \;,
\ee
each of which has
\be
\Delta H =  2^{N-1}\;, \qquad \Delta A = 2^{N-1}\;,
\ee
where we have denoted the eigenstates of $\sigma_{z}$ with
eigenvalues $+1$ and $-1$ by $|\!\uparrow\rangle$ and
$|\!\downarrow\rangle$ respectively.

For comparison,the maximum classical dispersion for the sum of
$2^{N-1}$ classical variables with factorized probability
distribution, each variable having maximum dispersion $1$, would be
$2^{(N-1)/2}$.

\vskip 0.2truein
\noindent\textbf{Parallel strategy CQC with exponentially enhanced
precision}: We start from one of the factorized $N$-qubit states
given above and apply one of the unitary operators $U = e^{- i\theta H}$
or $U = e^{- i\theta A}$ given above. In particular, we obtain
\beq  && |\psi _{H} (\theta)\rangle \;=\;e^{- i\theta
H}\; |\!\uparrow \uparrow \ldots \uparrow \uparrow \rangle \\
 &=& \cos (2^{N-1} \theta )|\!\uparrow \ldots  \uparrow \rangle
-i \sin ( 2^{N-1} \theta ) |\!\downarrow  \ldots
 \downarrow \rangle  \nonumber \; ;
\eeq
 \beq
 && |\psi _{A} (\theta)\rangle \;=\;e^{- i\theta
A}\; |\!\uparrow \uparrow \ldots \uparrow \uparrow \rangle \\
 &=& \cos (2^{N-1} \theta )|\!\uparrow  \ldots  \uparrow \rangle
+ \sin ( 2^{N-1} \theta ) |\! \downarrow \ldots \downarrow \rangle \;.
\nonumber
\eeq
 We may now measure the probability that all
qubits are in the up-state given by the expectation value of the
direct product of projection operators
\be
X=\bigotimes _{j=1} ^{N}\; \frac{1}{2}({\bf 1}+\sigma_{z}^{j}) \;.
\label{eq26}
\ee
We obtain, for example for the state $ |\psi _{H} (\theta)\rangle$
\be
\langle X\rangle = \frac{1}{2}\bigl[1+ \cos (\theta \, 2^{N})\bigr]\;,
\quad \Delta X = \frac{1}{2}\bigl|\sin (\theta \, 2^{N})\bigr| \;.
\label{eq27}
\ee
Hence, the quantum precision of estimating $\theta$
is given by
\be
\delta \theta = 2^{-N} \;,
\label{eq28}
\ee
 which achieves the best
allowed by the uncertainty relation since  $ \Delta H = 2^{N-1} $.
A similar precision is obtained by using the state $|\psi _{A}
(\theta)\rangle$. It is clear that we only needed to use the parallel
strategy $CQC$ (no entanglement in the probe-preparation or
measurement stage, but an entanglement-generating unitary operator
applied on the probes) to obtain this exponential enhancement of
precision.

\vskip 0.2truein
\noindent\textbf{A generalized Jaynes-Cummings model}: We propose a
quantum optics model of interaction of laser light with $N$ qubits
which could, in principle, be used to obtain exponentially enhanced
precision. We apply a unitary operator to the
infinite-dimensional vector space which implies an entanglement
generating `operation' on the reduced density operator for $N$
qubits.

Suppose we shine laser light of frequency  $\Omega \approx N \omega$
on an $N$-atom molecule each atom of which can be approximated to
be a two-level atom or qubit with level separation $\hbar \omega$ .
We assume that on irradiation,the molecule does not dissociate but
gets excited to a higher level in which each atom is excited to the
higher level. The second quantized interaction Hamiltonian is
assumed to be
\be
\hat H = \hat H_{0} + \hat H_{I}\;,\qquad
\hat H_{0}
=\frac{\omega}{2} \sum_{j=1} ^{N} \hat \sigma_{z} ^{j}
+\Omega \hat a^{\dagger} \hat a \;,
\ee
\be
\hat H_{I} = \frac{g}{2} \Bigl[ \hat a
\bigotimes _{j=1} ^{N}\bigl(\hat \sigma_{x} + i \hat \sigma_{y}
\bigr)^{j} + h.c. \Bigr] \;.
\ee
Here $h.c.$ denotes
Hermitian conjugate, $\hat a$ and $\hat a^{\dagger}$ denote annihilation
and creation operators for photons of frequency
$\Omega$, $\frac{1}{2}(\hat \sigma_{x} + i \hat \sigma_{y})^{j}$ denotes the
level raising operators for the $j$-th qubit and $g$ is a real
coupling constant. It is an exactly solvable Hamiltonian with
eigenvalues
\be
\lambda _{\pm} = \Omega (n+\frac{1}{2}) \pm \frac{\Omega_{1}}{2} \;,
\ee
where $n=0,1,2,\ldots$ and
 \be
\Omega_{1} = \sqrt{(\omega N - \Omega)^{2} + g^{2} (n+1) 2^{2 N} }
\;. \ee The corresponding eigenstates are, \be
\frac{1}{\sqrt{2}}\bigl(\alpha _{\pm}  | n \rangle|\phi_{0}\rangle +
\beta_{\pm} |n+1\rangle |\phi_{1}\rangle \bigr)\;, \ee
 \be
|\phi_{0}\rangle
 \equiv |\!\uparrow  \ldots \uparrow \rangle \;, \qquad
 |\phi_{1}\rangle  \equiv |\! \downarrow \ldots \downarrow \rangle \;,
\ee
\beq
\alpha_{\pm} & \equiv &\sqrt { 1\pm(\omega N - \Omega)/\Omega_{1}}\nonumber\\
\beta_{\pm} & \equiv & \pm \sqrt { 1\mp(\omega N -
\Omega)/\Omega_{1}}\;, \eeq
 and $\hat a^{\dagger} \hat a |n\rangle = n |n\rangle $.
Interesting physics about Rabi oscillations between up and down
qubit states may be read off from these equations even
off-resonance, i.e., $\Omega \neq N \omega $. For the present we
specialize to $\Omega = N \omega $ which implies $[\hat H_{0}, \hat
H_{I}]= 0$. We deduce that for $r=0$ and $r=1$
\beq
&&e^{- it \hat H}\, |n +r \rangle|\phi_{r}\rangle
 =e^{-it(n+1/2)N\omega}\nonumber \\
&&\times[ \cos (2^{N-1} \theta )|n + r\rangle |\phi_{r}\rangle \nonumber\\
&&\phantom{\times}
-i \sin ( 2^{N-1} \theta ) |n+1-r \rangle |\phi_{1-r}\rangle ] \;,
\eeq
 where $\theta = t g \sqrt {n+1}$. Thus the
reduced density operator for the $N$ qubits obtained by tracing over
the photon states undergoes the positivity and trace preserving
transformation \be \hat \rho (t)=\sum_{r =0,1 } \hat h_{r}\, \hat
\rho (0)\, \hat h_{r}^{\dag} \;,\qquad \hat h_{r}= e^{-i \theta H}\,
|\phi_{r}\rangle \langle\phi_{r}| \;, \ee where $H$ is the $N$-qubit
spin operator given by Eq.~(\ref{eq10}). For measurement of the
observable $X$ of Eq.~(\ref{eq26}) with $\langle  X\rangle = {\rm
tr} \; \hat \rho (t)  X $ we obtain exactly the same results on
precision as in Eqs.~(\ref{eq27}) and~(\ref{eq28}) of the previous
section. Investigation of practical implementability of this model
for achieving exponentially enhanced precision is an important
remaining task.

In conclusion, we have shown that the parameter estimation
associated with suitable entanglement-generating unitary operators
may lead to an exponential enhancement of accuracy over both
classical schemes and non-entangling quantum schemes. The exciting
thing is that unlike quantum computation, quantum metrology might
not face formidable problems of fighting decoherence if a suitable
interaction Hamiltonian can be found. We have shown that the
suggested strategy may be implemented in principle using a
generalized Jaynes-Cummings interaction.

\vskip 0.2truein
This work is funded in part by EPSRC grant EP/D500354/1.
SLB currently holds a Wolfson - Royal Society Research Merit award.

\end{document}